\documentclass[a4paper, 12pt]{article}
\usepackage[dvips]{graphics}
\usepackage[dvips]{epsfig}
\usepackage{makeidx}
\newcommand{\nc}{\newcommand}
\nc{\sgt}{\sigma_{\mbox{\scriptsize{tot}}}}
\nc{\sge}{\sigma_{\mbox{\scriptsize{el}}}}
\nc{\sges}{\sigma_{\mbox{\scriptsize{el}}}(s)}
\nc{\sgts}{\sigma_{\mbox{\scriptsize{tot}}}(s)}
\nc{\Sge}{\Sigma_{\mbox{\scriptsize{el}}}}
\nc{\slp}{$\left. d/dt \ln \mathrm{Im}\phi_+(s,t)\right|_{t=0}$}
\nc\be{\begin{equation}}
\nc\ee{\end{equation}}
\nc\bea{\begin{eqnarray}}
\nc\eea{\end{eqnarray}}
\nc\lb{\label} 
\nc {\re} {\mathop{\mathrm{Re}}} 
\nc {\im} {\mathop{\mathrm{Im}}}

\begin{document}
\title{Bounds in proton-proton elastic scattering at low momentum transfer}
\author{
A.~T. Bates and N.~H. Buttimore \\
\normalsize\emph{School of Mathematics, University of Dublin,} \\
\normalsize\emph{Trinity College, Dublin 2, Ireland}
\\ \normalsize \tt {atbates@maths.tcd.ie}, \tt{nhb@maths.tcd.ie}
\thanks{
        A.T.B. and N.H.B. are grateful to Enterprise Ireland for
	partial support under Scientific Research Project SC/96/778
        and International Programmes IC/1999/075, IC/2000/027.}
}
\maketitle
\begin{abstract}
We present a bound on the imaginary part of the single helicity-flip
amplitude for spin$\,1/2$-spin$\,1/2$ scattering at small momentum transfer.
The variational method of Lagrange multipliers
is employed to optimize the single-flip amplitude
using the values of $\sgt$, $\sge$ and diffraction slope
as equality constraints in addition to the
inequality constraints resulting from unitarity.
Such bounds provide important information related
to the determination of polarization of a proton beam.
In the case of elastic proton collisions the analyzing power
at small scattering angles
offers a method of measuring the polarization of a proton beam,
the accuracy of the polarization measurement depending on
the single helicity-flip amplitude.
The bound obtained on the imaginary part of the
single helicity-flip amplitude indicates that
the analyzing power for proton-proton collisions
in the Coulomb nuclear interference region
should take positive nonzero values at high energies.
\end{abstract}
\section{Introduction}
\setcounter{equation}{0}
The proton spin puzzle has intrigued experimentalists and theorists since the 
surprising result from the EMC experiment at CERN in 1988,
which
found a smaller than expected contribution to the spin of the proton 
from the component quarks.
The question, 
``where does the spin of the proton come from?''
remains unanswered~\cite{ansel, cheng}.
Recent data suggests a value of $\sim 20-30\%$ for the fraction of the spin 
carried by the {\em up}, {\em down} and {\em strange} quarks.
The contribution from the gluons and from the orbital angular momentum of the 
quarks and gluons is not completely known.
The Relativistic Heavy Ion Collider at Brookhaven National Laboratory plans 
to probe the proton structure using the deep inelastic scattering of protons
at high center-of-mass energies ($\sqrt s = 50-500$~GeV) and momentum transfers
($p_T \geq 10~\mbox{GeV}/c$)~\cite{bunce}. To measure the contribution of the gluons to 
the spin of the proton, with sufficient accuracy, a polarized proton beam
with a beam polarization error of $5\%$ is necessary.
\paragraph*{}
One method of measuring the polarization of a proton beam uses the
analyzing power in elastic proton collisions at small scattering angles.
The analyzing power $A_{\mbox\scriptsize{N}}$ for a proton and the
transverse single spin asymmetry ${\cal A}$ are related
through the expression
\be
A_{\mbox\scriptsize{N}}\,P= {\cal A}
\ee
where $P$ is the beam polarization; 
for $100\%$ beam polarization the
asymmetry and analyzing power are equal.
The beam polarization can be measured by counting the scatters with the beam
polarized up ($N^{\uparrow}$) and then down ($N^{\downarrow}$) in a
polarimeter with a known analyzing power
$A_{\mbox\scriptsize{N}}$:
\be
P = \frac{1}{A_{\mbox\scriptsize{N}}}
\left[\frac{N^{\uparrow} - N^{\downarrow}}
{N^{\uparrow} + N^{\downarrow}}\right] =
\frac{1}{A_{\mbox\scriptsize{N}}}\,{\cal A}\,.
\ee
The analyzing power $A_{\mbox\scriptsize{N}}$
expressed in terms of the $s$-channel helicity amplitudes is
\be
A_{\mbox\scriptsize{N}}\,\frac{d \sigma}{d t}=
-\frac{4 \pi}{s (s -4 m^2)}\,\im
\left[
\phi_5^{\ast}\,
\left(\phi_1 + \phi_2 + \phi_3 - \phi_4\right)
\right]\,.
\ee
        The reduced ratio, $r_5$,
        of the helicity single-flip to
        imaginary non-flip amplitude is defined as
\begin{equation}
       r_5  
	=
	{{m}\over{\sqrt{-t}}} \times
	{{\phi_5}
	\over
	{{\rm Im}\,\phi_+}}
	\, .
\end{equation}

        The analyzing power $A_{\mbox\scriptsize{N}}$
        for the CNI region can be written as
        follows, when the transverse total cross section spin
        difference is neglected~\cite{nhb99}:
\begin{equation}
        A_{\mbox\scriptsize{N}}
	=
        \frac{\sqrt{-t}}{m}\;
        {{(\kappa - 2 \im r_5 )(t_c/t) + 2 (\rho \im r_5 - \re r_5)}
	\over
	{(t_c/t)^2 - 2(\rho + \delta)(t_c/t) +1 + \rho^2 + \beta^2}}
\end{equation}
	where
$
        t_c =  - (8 \pi \alpha) / \sigma_{\rm tot} \, ,
$
$\delta$ is the Coulomb phase,
$\kappa + 1 = \mu = 2.7928$ is the magnetic moment of the proton, and
$\beta^2$ represents the forward hadronic double spinflip contributions
	expected to be negligible at high energies.
	Apart from the photon pole term,
	the $t$-dependences of helicity nonflip and flip
	electromagnetic and hadronic amplitudes
	due to form factor and nuclear slope effects
	are not expected to play a significant role
        in the amplitude ratios featuring in the asymmetry. Inside the CNI region
interference between the electromagnetic and hadronic amplitudes
is most prominent.
An important contribution to the maximum of
$A_{\mbox\scriptsize{N}}$, in the CNI region ($|t| < |t_c|$), 
comes from $\im r_5$ in the form of $\mu -1 - 2 \im r_5$.
At larger momentum transfers outside the
CNI region ($|t| > |t_c|$)  the analyzing power, containing hadronic terms only,
 is essentially
\begin{equation}
     A_{\mbox\scriptsize{N}}
	=
        \frac{\sqrt{-t}}{m}\;
        {{2\,(\rho \im r_5 - \re r_5)}
	\over
	{1 + \rho^2 + \beta^2}}
\end{equation}
and because the value of $\rho$ is expected to be no more
than about $10 \%$ at RHIC energies,
the analyzing power provides a good indication of the
value of $\re r_5$ just outside the CNI region.
The contribution of $\re r_5$ to the maximum of the analyzing power would be
 reasonably
well known.
A bound on $\im r_5$ which satisfies $\mu - 1 - 2 \im r_5 > 0$ ensures that
the maximum analyzing power in the CNI region is positive.
\paragraph*{}
The optimization technique of Lagrange multipliers,
extended by Einhorn and Blankenbecler~\cite{eb, mrs, eden} to include
equality and inequality constraints in the context of scattering theory,
is used to derive the bound
\be
|\im r_5|  
\leq
m\, \sqrt{g}\,
\left(\frac{36 \pi g \,\sge}{\sgt^2} -1 \right)^{1/2}\times h(t)
\ee
where
\be
h(t)= \frac{\left( 1 + g t(2 + 9 g t/8) \right)^{1/2}}{(1 + gt)}\,.
\ee
The partial wave expansions for the observables, $\sgt$, $\sge$,
and the hadronic slope parameter $g$ to be defined in Section 2,
are included as equality
constraints in the system, unitarity is presented as partial wave
inequality constraints and the modified imaginary single-flip amplitude, the
function to be optimized, is
input as the objective function.
Before proceeding to the derivation of the bound in Section 3 we first introduce the
$s$-channel helicity amplitudes, expanded as partial wave series
in the helicity representation. Errors on the derived bounds are discussed in
Section 4 followed by concluding remarks in Section 5.
\section{Helicity Amplitudes and Observables}
\setcounter{equation}{0}
For the elastic scattering of two protons at CM energy $\sqrt s$
and CM momentum $k= \sqrt{s -4m^2} /2$,
there are sixteen
helicity amplitudes in general, each a function of $s$ and $t$.
The number of independent amplitudes is reduced under 
the following relations~\cite{jw, ggmw};
\paragraph{Parity conservation}
\be
\left< \lambda_1^{\prime} \lambda_2^{\prime} | \,\phi\, | \lambda_1
\lambda_2 \right>
=
\left(-1\right)^{\mu -\lambda}
\left< -\lambda_1^{\prime} -\lambda_2^{\prime} | \,\phi\, | -\lambda_1 -\lambda_2\right>
\ee
\paragraph{Time reversal invariance}
\be
\left< \lambda_1^{\prime} \lambda_2^{\prime} | \,\phi\,  | \lambda_1 \lambda_2\right>
=
\left(-1\right)^{\mu -\lambda}
\left< \lambda_1 \lambda_2 | \,\phi\, |
\lambda_1^{\prime} \lambda_2^{\prime} \right>
\ee
\paragraph{Identical particle scattering}
\be
\left< \lambda_1^{\prime} \lambda_2^{\prime} | \,\phi\,  | \lambda_1 \lambda_2
\right>
=
\left(-1\right)^{\lambda-\mu}
\left< \lambda_2^{\prime} \lambda_1^{\prime} | \,\phi\, | \lambda_2
\lambda_1 \right>
\ee
where
$\lambda=\lambda_1-\lambda_2$,
$\mu=\lambda_1^{\prime} -\lambda_2^{\prime}$.

The sixteen helicity amplitudes 
reduce
to two non helicity-flip amplitudes $\phi_1$ and $\phi_3$,
two double helicity-flip amplitudes $\phi_2$ and $\phi_4$, and one
single helicity-flip amplitude $\phi_5$, with partial wave
expansions~\cite{jw, ggmw}:
\be
\phi_1 \left(s,t\right) = \left<+ + | \phi | + +\right> =
\frac{\sqrt s}{2k}
\sum_{J}^{} \left(2J+1\right)
\left(f_{0}^J(s)+f_{11}^J(s)\right)
d_{00}^J\left(\theta\right)
\lb{eq:phi1} 
\ee
\be
\phi_3 \left(s,t\right) =\left<+ - | \phi | + -\right>
=\frac{\sqrt s}{2k}
\sum_{J}^{} \left(2J+1\right)
\left(f_{1}^J(s)+f_{22}^J(s)\right)
d_{11}^J\left(\theta\right)
\lb{eq:phi3}
\ee
\be
\phi_2 \left(s,t\right)=\left<+ + | \phi | - -\right>  
=\frac{\sqrt s}{2k}
\sum_{J}^{} \left(2J+1\right)
\left(f_{11}^J(s)-f_{0}^J(s)\right)
d_{00}^J\left(\theta\right)
\lb{eq:phi2}
\ee
\be
\phi_4 \left(s,t\right)=\left<+ - | \phi | - +\right> 
=\frac{\sqrt s}{2k}
\sum_{J}^{} \left(2J+1\right)
\left(f_{22}^J(s)-f_{1}^J(s)\right)\!
d_{1-1}^J\left(\theta\right)
\lb{eq:phi4}
\ee
\be
\phi_5 \left(s,t\right)=
\left<+ + | \phi | + -\right> 
=\frac{\sqrt s}{2k} \sum_{J}^{} \left(2J+1\right)
f_{21}^J(s)\,
d_{10}^J\left(\theta\right)
\lb{eq:phi5}
\ee
where $f_i^J \left(i=0, 1, 11, 22, 21\right)$ denote $s$-channel partial
wave amplitudes, $\im f_i^J = a_i^J$, $\re f_i^J = b_i^J$ and 
$z=\cos \theta=1 + t/2k^2$.
In the Coulomb Nuclear Interference (CNI) region, $t \approx -0.0012
(\mbox{GeV}/c)^2$, it is convenient to express the five
helicity amplitudes in terms of Jacobi polynomials.
To define the $d_{\lambda\,\mu}^J(\theta)$ function in terms of Jacobi
polynomials it is suitable to separate the space of $\lambda$ and $\mu$ into
four regions A, B, C, D~\cite{andrwgun}.
In region A, where $\lambda +\mu \geq 0$ and $\lambda -\mu \geq 0$,
the relation is
\be
d_{\lambda\,\mu}^J(\theta)=
\left(
\sqrt{
\frac{(J + \lambda)!\,(J -\lambda)!}
{(J + \mu)!\,(J -\mu)!}
}\,
\left(
\frac{1+z}{2}
\right)^{\frac{(\lambda + \mu)}{2}}\,
\left(
\frac{1-z}
{2}
\right)^{\frac{(\lambda - \mu)}{2}}
\right) \times
P_{J - \lambda}^{(\lambda - \mu,\,\lambda + \mu)} (z)\,,
\lb{eq:dj}
\ee
and $J - \lambda = 0, 1, 2, \cdots$.
Equivalent forms in the other regions are obtained by use of symmetry relations
~\cite{ggmw, andrwgun}. \newline
In region B, where $\lambda +\mu \geq 0$ and $\lambda -\mu \leq 0$, use
\be
d_{\lambda\,\mu}^J(\theta) = (-1)^{\lambda - \mu}\,d_{\mu\,\lambda}^J(\theta)
\lb{eq:symb}\,.
\ee
In region C, where $\lambda +\mu \leq 0$ and $\lambda -\mu \leq 0$, use
\be
d_{\lambda\,\mu}^J(\theta) = (-1)^{\lambda - \mu}\,d_{-\lambda\,-\mu}^J(\theta)
\lb{eq:symc}\,.
\ee
In region D, where $\lambda +\mu \leq 0$ and $\lambda -\mu \geq 0$, use
\be
d_{\lambda\,\mu}^J(\theta) = d_{-\mu\,-\lambda}^J(\theta)
\lb{eq:symd}\,.
\ee
Expressing the $d_{\lambda\,\mu}^J(\theta)$ functions in terms of
Jacobi polynomials the five independent helicity amplitudes
can be written as
\bea
\phi_1(s,t) &=&
\frac{\sqrt s}{2k}
\sum_{J}^{} \left(2J+1\right)
\left(f_{0}^J(s)+f_{11}^J(s)\right)\,
P_{J}^{(0,\,0)}(z)
\lb{eq:phi1jp}
\\
\phi_2(s,t) &=& \frac{\sqrt s}{2k}
\sum_{J}^{} \left(2J+1\right)
\left(f_{11}^J(s)-f_{0}^J(s)\right)
P_{J}^{(0,\,0)}(z)
\lb{eq:phi2jp}\,
\\
\phi_3(s,t) &=& \frac{\sqrt s \,(1+z)}{4k}
\sum_{J}^{} \left(2J+1\right)
\left(f_{1}^J(s)+f_{22}^J(s)\right)
P_{J-1}^{(0,\,2)}(z)
\lb{eq:phi3jp}
\\
\phi_4(s,t) &=& \frac{\sqrt s \,(1 - z)}{4k}
\sum_{J}^{} \left(2J+1\right)
\left(f_{22}^J(s)-f_{1}^J(s)\right)
P_{J-1}^{(2,\,0)}(z)
\lb{eq:phi4jp}
\\
\phi_5 (s,t) &=& \frac{\sqrt s\,\sqrt{1-z^2} }{4k}
\sum_{J}^{} \left(2J+1\right)\, \sqrt{ \frac{J+1}{J} }\,
f_{21}^J(s)\,
P_{J-1}^{(1,\,1)}(z)\
\lb{eq:phi5jp}
\eea
where $z=\cos \theta = 1 + t/(2 k^2)$. The Jacobi polynomials have the 
properties~\cite{bell}
\be
P_n^{(\alpha, \beta)}(1)=
\frac{\Gamma(\alpha + n +1)}{\Gamma(\alpha +1) n!}\,,
\lb{eq:jpz1}
\ee
\be
\frac{d^m}{d z^m}\, P_n^{(\alpha, \beta)}(z) =
2^{-m} \frac{\Gamma(m + n + \alpha+ \beta +1)}
{\Gamma(n + \alpha+ \beta +1)} \,
P_{n-m}^{(\alpha+m, \beta+m)}(z)\,.
\lb{eq:jpderiv}
\ee
\subsection{Observables}
In proton-proton elastic scattering the spin observables
can be written in terms of the five helicity amplitudes, $\phi_1, \ldots,
\phi_5$. The observables are vital to the optimization,  since
each observable can be included as an equality constraint in the
optimized system. In the derivation of the bound we use three equality constraints,
$\sgt$, $\sge$ and $g$, and two inequality constraints related to unitarity. 
\subsubsection{Total Cross Section}
The first equality constraint uses the total cross section.
The optical theorem,
\be
\left. \im \phi_+(s,t) \right|_{t=0} =
\frac{k\, \sqrt s}{2 \pi}\, \sgts\,,
\lb{eq:opthm}
\ee
is used to express the total cross section as a partial wave expansion
given by
\be
\sgts 
= 
\frac{\pi}{k^2}
\sum_{J}^{} \left(2J+1\right)
\left\{  
\left(a_{0}^J+a_{11}^J\right)\,
P_{J}^{(0,\,0)}(1)
+\left(a_{1}^J+a_{22}^J\right)\,
P_{J-1}^{(0,\,2)}(1) \right\}
\ee
where $\im \phi_+(s,t) = ( \im \phi_1(s,t) + \im \phi_3(s,t) )/2$ is the
imaginary spin average helicity non-flip amplitude. 
Using property~(\ref{eq:jpz1}) the 
normalized dimensionless total cross section can be expressed as a 
partial wave expansion:
\be
A_0 = 
\frac{\pi}{k^2}
\sum_{J}^{} \left(2J+1\right)
\left\{
a_{0}^J(s)+a_{1}^J(s)+ a_{11}^J(s)+a_{22}^J(s)
\right\}
\lb{eq:sgt}
\ee
where $A_0=(k^2/\pi)\; \sgt$.
\subsubsection{Slope of the Imaginary Non-Flip Amplitude}
The slope of the imaginary non-flip amplitude has been used in bounds for other
spin dependent elastic collisions~\cite{martin}. We find it convenient to use
the imaginary part of the spin-averaged amplitude at a particular value
of $t$.
The second equality constraint employs the 
imaginary spin average non-flip amplitude
at a particular $t$ inside the Coulomb Nuclear Interference region, written
as a Taylor expansion:
\be
\im \phi_+(s,t)
 \approx 
\im \phi_+(s,0) +
t \left(
\left. \frac{d}{dt} \im \phi_+(s,t)
\right)
\right|_{t = 0}\,,
\lb{eq:tayphi+}
\ee
where $|t|$ is sufficiently small so that inclusion of the linear term in the
Taylor expansion is an accurate approximation.
Use of properties~(\ref{eq:jpz1}) and ~(\ref{eq:jpderiv}) leads to
the partial wave expansion for the imaginary non-flip amplitude:
\be
\im \phi_+(s,t)=
\frac{\sqrt s}{4 k}
\sum_{J}^{} \left(2J+1\right)
\left\{
a_{0}^J+a_{1}^J+ a_{11}^J+a_{22}^J
\right\}\,
\left( 1 - \frac{\zeta}{4}\, J(J+1) \right)\
\lb{eq:Bslp}
\ee
where $\zeta= -t/k^2$.
The logarithmic derivative of the imaginary spin average non-flip amplitude,
\be
g = \frac{d}{dt} \ln \left.\im \phi_+(s,t)\right|_{t = 0} =
\frac{1}{\im \phi_+(s,0)}
\left.
\left(\frac{d}{dt} \im \phi_+(s,t)\right)
\right|_{t = 0}
\lb{eqn:g1}\,,
\ee
can be expressed as 
\be
\left.
\left(\frac{d}{dt} \im \phi_+(s,t)\right)
\right|_{t = 0}
= g\,\im \phi_+(s,0) \,.
\ee
The Taylor expansion for $\im \phi_+(s,t)$, given by
Equation~(\ref{eq:tayphi+}), can thus be
written as
\be
\im \phi_+(s,t)= 
\im \phi_+(s,0)\, \left\{ 1 - (-t\,g) \right\}  \,.
\ee
\subsubsection{Elastic Cross Section}
The third equality relates to the elastic cross section,
expressed as a partial wave expansion by
integrating the differential cross section over momentum transfer $t$:
\be
\sges =
\int_{- 4 k^2}^{0} dt \frac{d \sigma(s,t)}{dt}\,.
\ee
Expressing the differential cross section in terms of helicity amplitudes
allows us to 
write the elastic cross section as
\be
\sges =
\frac{\pi}{2 k^2\,s}\,\int_{- 4 k^2}^{0} dt
\left\{
|\phi_1|^2 +  |\phi_2|^2 + |\phi_3|^2  
 +  |\phi_4|^2 +
4 |\phi_5|^2
\right\}\,.
\ee
Using the expression
\be
t = -2 k^2 (1 -z)\,,
\ee
the $t$ variable can be replaced with the $z$ variable,
where $t$ is the momentum transfer, $k$ is the center-of-mass three-momentum
and $z= \cos \theta$. The elastic cross section expressed as an integral
over $z$ becomes
\be
\sges
=
\frac{\pi}{s}\,\int_{-1}^{+1} dz
\left\{
|\phi_1|^2 +  |\phi_2|^2 + |\phi_3|^2  
+ |\phi_4|^2 +
4 |\phi_5|^2
\right\}\,.
\lb{eq:sigelz}
\ee
To express the elastic cross section as a partial wave expansion,
the integrals
$$\int_{-1}^{+1} dz |\phi_i(s,t)|^2$$ are calculated, where
$i= 1, \cdots, 5$.
The integration formula~\cite{bateman}
\be
\int_{-1}^{+1}
(1 - z)^{\alpha}(1 + z)^{\beta}\,
P_n^{(\alpha, \beta)}(z)\,
P_m^{(\alpha, \beta)}(z)\,dz
=
\left\{
\begin{array}{cc}
&\hspace*{-30mm}0 \;\;\;m \not= n \\
&\hspace*{-5mm} \frac{2^{\alpha+\beta+1}\,\Gamma(\alpha + n +1)\, \Gamma(\beta + n +1)}
{n!\,(\alpha+ \beta + 2n +1)\,\Gamma(\alpha + \beta+ n +1)}\,
\delta_{m,\,n}\,,
\end{array}
\right.
\ee
may be used to find the integral
\be
\int_{-1}^{+1}
d^J_{\lambda \,\mu}(\theta)\,d^M_{\lambda \,\mu}(\theta)\,dz=
\frac{2}{2J+1}\, \delta_{M,\,J}\,,
\lb{eq:intdj}
\ee
leading to a partial wave expansion for the normalized dimensionless
elastic cross section, defined as $\Sge= (k^2/\pi)\;\sge$:
\be
\Sge(s)=
\sum_{J}^{} \left(2J+1\right)
\left\{
\left|f_{0}^J\right|^2 +
\left|f_{1}^J\right|^2 +
\left|f_{11}^J\right|^2                                
+\left|f_{22}^J\right|^2 +
2 \left|f_{21}^J\right|^2
\right\}\,.
\lb{eq:Sig}
\ee
\subsection{Imaginary Single-Flip Amplitude}
The imaginary single helicity-flip amplitude, modified by a kinematical factor,
is the objective function in the system. 
Before optimization we must first express the single-flip amplitude
in a suitable form. The
imaginary amplitude,
\be
\im \phi_5 \left(s,t\right)  =  \frac{\sqrt s}{2k}
\sum_{J}^{} \left(2J+1\right)
a_{21}^J\,
d_{10}^J\left(\theta\right)\,,
\lb{eq:phi51}
\ee
written in terms of Jacobi polynomials and normalized by a kinematical factor,
becomes
\be
\im \tilde \phi_5 =
\frac{\im \phi_5 (s,t)}{(1-z^2)^{1/2}} = \frac{\sqrt s}{4k}
\sum_{J}^{} \left(2J+1\right)\, \sqrt{ \frac{J+1}{J} }\,
a_{21}^J\,
P_{J-1}^{(1,\,1)}(z)\,.
\lb{eq:phi5jp1}
\ee
In the CNI region, the Jacobi polynomial $P_{J-1}^{(1, 1)}(z)$
expanded as a Taylor series is
\be
P_{J-1}^{(1, 1)}(z) \approx
\left.P_{J-1}^{(1, 1)}(z)\right|_{z = 1} -
\frac{\zeta}{2}\,
\left.\left(\frac{d}{dz}\,P_{J-1}^{(1, 1)}(z)\right)\right|_{z=1}
\ee
where $z=1+ t/(2k^2)$ and $\zeta = - t/k^2$.
Using properties ~(\ref{eq:jpz1})
and (\ref{eq:jpderiv}), the Taylor series for $P_{J-1}^{(1, 1)}(z)$ about
$z =1$ is
\be
P_{J-1}^{(1, 1)}(z) \approx
J \left( 1 - \frac{\zeta}{8}\, \left[J(J+1) - 2 \right] \right)\,,
\ee
and thus, inside the CNI region, 
\be
\im \tilde \phi_5 \approx
\frac{\sqrt s}{4k}
\sum_{J}^{} \left(2J+1\right)\, \sqrt{ \frac{J+1}{J} }\,
J \left( 1 - \frac{\zeta}{8}\, \left[J(J+1) - 2 \right] \right)\,
a_{21}^J\,.
\lb{eq:phi5norm}
\ee
At small collision angles,
\be
\frac{m}{\sqrt{-t}} \approx
\frac{m}{k\,(1 -z^2)^{1/2}}\,,
\ee 
the ratio $\im r_5= m \im \phi_5 /(\sqrt{-t} \im \phi_+)$ is 
$$
\im r_5 =
\frac{m}{k}\,\frac{\im \tilde \phi_5}{\im\phi_+(s,t)}\,.
\lb{eq:imr5}
$$
\subsection{Unitarity}
The partial wave amplitudes obey the following unitarity 
inequalities~\cite{mu1}
\bea
U_1^J &=& a_{0}^J - |f_{0}^J|^2 \geq 0
\lb{eq:pwu1} \\
U_2^J &=& a_{1}^J - |f_{1}^J|^2 \geq 0 
\lb{eq:pwu2}\\
V_1^J &=& a_{11}^J - |f_{11}^J|^2 - |f_{21}^J|^2 \geq 0
\lb{eq:pwu3} \\
V_2^J &=& a_{22}^J - |f_{22}^J|^2 - |f_{21}^J|^2 \geq 0
\lb{eq:pwu4}
\eea
where $f_i^J \left(i=0, 1, 11, 22, 21\right)$ denote the $s$-channel partial
wave amplitudes, $\im f_i^J = a_i^J$ and $\re f_i^J = b_i^J$.
Combining Equation~(\ref{eq:pwu1}) with Equation~(\ref{eq:pwu2}) leads to
the inequality $X^J=U_1^J+U_2^J$ where
\be
X^J=  a_{0}^J + a_{1}^J - |f_{0}^J|^2 - |f_{1}^J|^2  \geq 0
\lb{eq:W}
\ee
and the inequality $W^J=V_1^J+V_2^J$ follows from 
the combination of Equation~(\ref{eq:pwu3}) and Equation~(\ref{eq:pwu4})
where
\be
W^J =
a_{11}^J + a_{22}^J - |f_{11}^J|^2 - |f_{22}^J|^2 - 2 |f_{21}^J|^2 \geq 0\,.
\lb{eq:X}
\ee
\paragraph*{}
For the elastic scattering of spin 0 on spin 1/2 particles 
there are two independent helicity amplitudes, a flip and a non-flip amplitude,
with partial wave expansions whose partial wave amplitudes obey unitarity
relations similar to relations (\ref{eq:pwu1}) and (\ref{eq:pwu2}).
The unitarity relations (\ref{eq:pwu3}) and (\ref{eq:pwu4}) are
characteristic of spin 1/2 - spin 1/2 scattering, the $f_{21}^J$ term
coming from the single helicity-flip amplitude $\phi_5$.
\section{Optimization}
\setcounter{equation}{0}
Equipped with partial wave expansions for the observables and partial wave
inequality relations, representing unitarity, we are in a position to optimize
the modified helicity single-flip amplitude $\im \tilde \phi_5$.
We follow the variational technique of Einhorn and Blankenbecler~\cite{eb}
by constructing a Lagrangian consisting of an objective function and 
a set of equality and inequality constraints.
We use the
full set of constraints, $\sgt$, $\sge$, $g$ and unitarity, although a bound
on $\im \tilde \phi_5$, with fewer constraints, can be derived~\cite{atb}.
\subsection{Lagrange Formalism}
The normalized dimensionless total cross section $A_0$,
expressed as an equality constraint, is included in the Lagrange function
along with the normalized dimensionless elastic
cross section $\Sge$, written as an equality constraint, the imaginary
spin average non helicity-flip amplitude $\im \phi_+(s, t)$ at a fixed
small $|t|$ value,
also expressed as an equality
constraint, and
the partial wave unitarity relations written as inequality constraints.
The modified single helicity-flip amplitude $\im \tilde{\phi_5}$
is introduced as the objective function in the Lagrange function:
\bea
\cal{L} & = & \im \tilde{\phi_5} + 
 \alpha \left[
 A_0 - \sum_{J}^{} \left(2J+1\right)
\left\{
a_{0}^J+a_{1}^J+ a_{11}^J+a_{22}^J
\right\}
\right]
 \nonumber \\
& &+ \beta
\left[ 
\Sigma_{\mbox{\scriptsize{el}}} -
\sum_{J}^{} (2J+1)
\left(
|f_{0}^J|^2 + |f_{1}^J|^2 + |f_{11}^J|^2 +
|f_{22}^J|^2 +2 |f_{21}^J|^2 
\right)
\right] \nonumber \\
& &
\!\!\!\!\!\!\!\!\!\!\!\!\!\!\!\!\!\!+\,\gamma
\left[ 
\im \phi_+ -
\frac{\sqrt s}{4 k}
\sum_{J}^{} \left(2J+1\right)
\left\{
a_{0}^J+a_{1}^J+ a_{11}^J+a_{22}^J
\right\}\,
\left( 1 - \frac{\zeta}{4}\, J(J+1) \right)
\right] \nonumber \\
& &
+\sum_{J}^{}(2J+1)\mu_{J}
\left(
a_{11}^J + a_{22}^J - |f_{11}^J|^2 - |f_{22}^J|^2 - 2 |f_{21}^J|^2
\right) 
\nonumber \\
& &
+\sum_{J}^{}(2J+1) \lambda_{J}
\left( a_{0}^J + a_{1}^J - |f_{0}^J|^2 - |f_{1}^J|^2 \right) 
\lb{eq:lagr4c1}
\eea
where $\alpha$, $\beta$ and $\gamma$ are equality multipliers.
The inequality
multipliers, $\lambda_{J}$ and $\mu_{J}$, are by definition non-negative
and $\zeta= -t/(k^2)$.
In the high energy or large $J$ limit only the leading
order $J$ terms are included and
the Lagrange function of Equation~(\ref{eq:lagr4c1}) becomes
\bea
\cal{L} & = & \im \tilde{\phi_5} + 
 \alpha \left[
 A_0 - 2 \sum_{J}^{} J
\left\{
a_{0}^J+a_{1}^J+ a_{11}^J+a_{22}^J
\right\}
\right]
\nonumber \\
& &+\beta
\left[ 
\Sigma_{\mbox{\scriptsize{el}}} -
2\sum_{J}^{} J
\left(
|f_{0}^J|^2 + |f_{1}^J|^2 + |f_{11}^J|^2 +
|f_{22}^J|^2 +2 |f_{21}^J|^2 
\right)
\right] \nonumber \\
& &
+\gamma
\left[ 
\im \phi_+ -
\sum_{J}^{} J
\left\{
a_{0}^J+a_{1}^J+ a_{11}^J+a_{22}^J
\right\}\,
\left( 1 - \frac{\zeta}{4}\,J^2\right)
\right] \nonumber \\
& &
+2 \sum_{J}^{} J\mu_{J}
\left(
a_{11}^J + a_{22}^J - |f_{11}^J|^2 - |f_{22}^J|^2 - 2 |f_{21}^J|^2
\right) 
\nonumber \\
& &
+2 \sum_{J}^{}J\lambda_{J}
\left( a_{0}^J + a_{1}^J - |f_{0}^J|^2 - |f_{1}^J|^2 \right) 
\lb{eq:lagr4c2}
\eea
and
\be
\im {\tilde\phi_5} \approx
\sum_{J}^{}J^2 \,
\left( 1 - \frac{\zeta}{8}\,J^2 \right)\,
a_{21}^J\,.
\lb{eq:imf5_2}
\ee
The system is optimized by taking first and second derivatives 
with respect to the real
and imaginary partial wave amplitudes, $b_i^J$ and $a_i^J$.
This gives the optimized set of partial waves, at some fixed $t$ in the
CNI region;
\bea
b_i^J &=& 0 \;\forall i\,, \lb{optpw10}
\eea
\be
a_0^J = a_1^J =
\frac{r_1 + r_2 \left(1 -\frac{\zeta}{4}\,J^2\right) +\tilde\lambda_J}
{1 + 2\tilde\lambda_J}\,,
\lb{eq:optpw4c0}
\ee
\bea
a_{11}^J = a_{22}^J =
\frac{r_1 + r_2 \left(1 -\frac{\zeta}{4}\,J^2\right) +\tilde\mu_J}
{1 + 2\tilde\mu_J}
\lb{eq:optpw4c11}
\eea
and
\be
a_{21}^J =
\frac{\frac{J}{8 \beta}
\left(
1 - \frac{\zeta}{8}\,J^2
\right)
}
{1 + 2 \tilde\mu_J}
\lb{eq:optpw4c21}
\ee
where $\tilde \lambda_J =  \lambda_J  / 2 \beta$, 
$\tilde \mu_J =  \mu_J  / 2 \beta$,
$r_1= -\alpha/(2 \beta)$, $r_2= -\gamma/(4 \beta)$
and $\beta > 0$ for a maximum (or $\beta <0$ for a minimum).
\subsection{Unitarity Classes}
Optimization under the four constraints imposes the following condition:
\be
b_i^J = 0 \;\forall i \Longrightarrow f_i^J = a_i^J + b_i^J = a_i^J\,,
\ee 
that is, there is no contribution from the real partial wave amplitudes.
The imaginary partial wave amplitudes therefore
obey the following unitarity inequalities:
\be
X^J=  a_{0}^J - a_{0}^J{}^2 \geq 0
\lb{eq:pw1}
\ee
\be
W^J =
a_{11}^J - a_{11}^J{}^2 - a_{21}^J{}^2 \geq 0
\lb{eq:pw2}
\ee
where 
$
a_0^J = a_1^J\;\;\;\mbox{and}\;\;\;a_{11}^J = a_{22}^J
$.
It is natural to divide the partial waves into two classes,
one with contributions from the interior unitarity class $I$ and the other
with contributions from the boundary unitarity class $B$.
For the $X^J$ unitarity inequality the interior and boundary unitarity
classes are defined as
\bea
I^X \equiv \left\{ J \,|\, X_J > 0,\,\tilde \lambda_J = 0\right\} 
\,,& &
B^X \equiv \left\{ J \,|\, X_J = 0,\,\tilde \lambda_J \ge0\right\}\,.
\eea
Likewise for the $W^J$ unitarity inequality the interior and boundary
unitarity
classes are 
\bea
I^W \equiv\left\{ J \,|\, W_J > 0,\,\tilde \mu_J = 0\right\} 
\,,& &
B^W \equiv \left\{ J \,|\, W_J = 0,\,\tilde \mu_J \ge0\right\} \,.
\eea
\subsubsection{$I^X$ and $B^X$ Unitarity Classes}
The interior unitarity class,
\be
I^X \equiv
\left\{ J \,|\,X^J = a_{0}^J - a_{0}^J{}^2 > 0,\,\tilde \lambda_J =
0\right\}\,,
\ee
under the four constraints,
is expressed as
\be
I^X \equiv
\left\{ J \,|\,0 < a_0^J < 1,\,\tilde \lambda_J = 0\right\}\,.
\ee
Equation~(\ref{eq:optpw4c0}) with $\tilde \lambda_J$ set to zero
enables us to write
the imaginary partial wave amplitude $a_0^J$, in the interior
unitarity class, as
\be
a_0^J= r_1+ r_2 \left(1 -\frac{\zeta}{4}\,J^2\right)\,.
\ee
The constraint $0 < a_0^J <$ 1 restricts the values of the equality
multipliers, $r_1$ and $r_2$, to $0 < r_1+r_2 < 1$ and $r_2 >0$. The number of
partial waves $J$ are thus limited to
\be
0 \leq  J^2 < \frac{4}{\zeta}\,\left(1+ \frac{r_1}{r_2} \right)\,.
\ee
\paragraph{}
The boundary unitarity class $B^X$ splits into two sub-classes, $B^{X_0}$
and $B^{X_1}$:
\bea
&\!\! \longrightarrow\!\! & B^{X_0}\equiv \left\{ J \,|\,a_0^J =0,\,
\tilde \lambda_J \geq 0\right\} \nonumber\\
B^X \equiv\left\{ J \,|\, X_J = a_{0}^J - a_{0}^J{}^2 = 0,\,\tilde \lambda_J \geq 0\right\}
&  &  \nonumber\\
&\!\! \longrightarrow \!\!& B^{X_1} \equiv \left\{ J \,|\,a_0^J =1,\,
\tilde \lambda_J \geq 0\right\} \nonumber \,.   \\
& & 
\eea
In the boundary unitarity class $B^{X_0}$ the imaginary partial wave
amplitude $a_0^J$ is equal to zero and from Equation~(\ref{eq:optpw4c0})
the inequality multiplier $\tilde \lambda_J$ is given by
\be
\tilde \lambda_J=
-(r_1+r_2) + r_2 \,\frac{\zeta}{4}\,J^2 \geq 0\,.
\ee
The $B^{W_0}$ class begins at $J^2 = M_1^2= 4/\zeta\,(1+ r_1/r_2)$, and
for $J \geq M_1+1$, with $0 < r_1+r_2 < 1$ and $r_2>0$,
the inequality multiplier
$\tilde \lambda_J$ is positive. Therefore the boundary unitarity class
$B^{X_0}$ is non-empty for $J \geq M_1+1$ but with $a_0^J=0$, for all $J$
in this unitarity class,
there are no contributions to the observables from this unitarity class.
The imaginary partial wave amplitude $a_0^J$ is equal to unity in the
boundary unitarity class $B^{X_1}$ and from  Equation~(\ref{eq:optpw4c0})
the inequality multiplier $\tilde \lambda_J$ is given by
\be
\tilde \lambda_J=
(r_1+r_2) - 1 - r_2 \,\frac{\zeta}{4}\,J^2\,.
\ee
By definition $\tilde \lambda_J \geq 0$ and the value of $J$, in the
boundary unitarity class, is limited to
\be
J^2 \leq \frac{4}{\zeta} \,\left(\frac{(r_1+r_2)-1}{r_2}\right)
\ee
but with  $0 < r_1+r_2 < 1$ and $r_2>0$, $J^2$ is negative, or $J$ is complex
and therefore the boundary unitarity class $B^{X_1}$ is empty.
\paragraph{}
In summary, the unitarity classes, $I^X$ and $B^{X_0}$, are non-empty and the
unitarity class $B^{X_1}$ is empty;
\be
I^X \equiv \left\{J | 0 < a_0^J < 1,\; 0 < J \leq M_1\right\},
\ee
\be
B^{X_0}  \equiv \left\{J | a_0^J = 0,\; M_1+1 \leq J \leq M_2\right\}
\ee
where $M_1={\tt Floor}\left[\sqrt{4/\zeta\,(1+ r_1/r_2)}\,\right]$,
$M_2={\tt Floor}\left[\sqrt{8/\zeta}\,\right]$ and
$\zeta=-t/k^2$. The {\tt Floor[x]} function gives the greatest integer less
than or equal to {\tt x}.

\subsubsection{$I^W$ and $B^W$ Unitarity Classes}
The interior unitarity class $I^W$ under the optimization becomes
\be
I^W \equiv
\left\{ J \,|\,W^J=a_{11}^J - a_{11}^J{}^2 - a_{21}^J{}^2 > 0\,,\,
\tilde \mu_J = 0\right\},
\lb{eq:IX}
\ee
Substituting Equations~(\ref{eq:optpw4c11}) and (\ref{eq:optpw4c21}),
with $\tilde \mu_J = 0$, into
the interior constraint $a_{11}^J - a_{11}^J{}^2 - a_{21}^J{}^2 > 0$
leads to the equation;
\be
f_2(J)=\tilde a_1 + \tilde a_2\,J^2 + \tilde a_3\,J^4 + \tilde a_4\, J^6 > 0
\ee
where 
$\tilde a_1=(r_1+r_2)\,(1 - (r_1+r_2)),\,
\tilde a_2=r_2\zeta\,( 2(r_1+r_2) - 1)/4- 1/(64 \beta^2),\,
\tilde a_3=\zeta/(256 \beta^2)-r_2^2 \zeta^2/16,\,
\tilde a_4= -\zeta^2/(64\beta)^2$,
and only positive $J$ solutions are allowed.
The solution is of the form
\be
0 < J^2 < \eta_{2}^2 \,\frac{4}{\zeta}\,\left(1+ \frac{r_1}{r_2} \right)
=\eta_2^2\,M_1^2
\ee
where $\eta_2\sim 1$.
The boundary unitarity class $B^W$ is written as
\be
B^W \equiv
\left\{ J \,|\,W^J=a_{11}^J - a_{11}^J{}^2 - a_{21}^J{}^2 = 0,\,
\tilde \mu_J \geq 0\right\}.
\ee
The constraint $a_{11}^J - a_{11}^J{}^2 - a_{21}^J{}^2 = 0$ can be
written as a quadratic equation:
\be
\tilde \mu_J^2 + \tilde \mu_J + f_2(J)=0
\ee
where
\be
f_2(J)= \tilde a_1 + \tilde a_2\,J^2 + \tilde a_3\,J^4 + \tilde a_4\, J^6\,.
\ee
The solutions are
\be
\tilde \mu_J =
\frac{1}{2}\,
\left\{
\pm \sqrt{ 1 - 4 f_2(J) } -1 \right\} \,.
\ee
The function $f_2(J)$ is negative for $J > M_1 =
{\tt Floor}[4/\zeta\,\left(1+ r_1/r_2 \right)]$ and
consequently $\tilde \mu_J$ is positive for such $J$ values.
By definition $\tilde \mu_J \geq 0$, therefore the positive solution is chosen;
\be
\tilde \mu_J =
\frac{1}{2}\,
\left\{
\sqrt{ 1 - 4 f_2(J) } -1 \right\} \,.
\ee
To summarize, both the classes, $I^W$ and $B^W$, are
non-empty:
\be
I^W \equiv \left\{J | a_{11}^J - a_{11}^J{}^2- a_{21}^J{}^2 > 0,
\; 0 \leq J \leq M_1\right\}\,,
\ee
\be
B^{W}  \equiv \left\{J | a_{11}^J - a_{11}^J{}^2- a_{21}^J{}^2 =0,\;
M_1+1 \leq J \leq M_2\right\}\,,
\ee
with $\eta_2=1$, where
$
M_1={\tt Floor}
\left[\sqrt{4/\zeta\,\left(1+ r_1/r_2 \right)}\,\right]
$, 
$
M_2={\tt Floor}
\left[\sqrt{8/\zeta}\,\right]
$
and
$\zeta=-t/k^2$.
It is important to notice that with $\eta_2=1$ both interior unitarity classes,
$I^W$ and $I^X$, are non-empty over the same region, $J \in [0, M_1]$.
Similarly the boundary unitarity classes,
$B^{X_0}$ and $B^W$, are non-empty over the same region,
$ M_1+1 \leq J \leq M_2$.
In other
words there is no mixing of unitarity classes, all classes are either
interior unitarity classes, $I\equiv I^X \cup I^W$,
or boundary unitarity classes, $B\equiv B^X \cup B^W$, for a given $J$.
\paragraph{}
\subsection{Solution of Interior Unitarity Class}
Consider the set of interior unitarity classes,
$I \equiv I^X \cup I^W$.
The inequality multipliers, $\tilde \lambda_J$ and $\tilde \mu_J$,
in the interior region are equal to zero.
The imaginary partial wave amplitudes are therefore written as
\be
a_k^J= r_1+ r_2 \left(1 -\frac{\zeta}{4}\,J^2\right)
\lb{eq:ak1}
\ee
and
\be
a_{21}^J =
\frac{J}{8 \beta}
\left(
1 - \frac{\zeta}{8}\,J^2
\right)\,,
\lb{eq:ak21}
\ee
$k=0, 1, 11, 22$, with $0 \leq J \leq M_1$, where
$M_1={\tt Floor}\left[\sqrt{4/\zeta\,(1 +  r_1/r_2)}\,\right]$
is the maximum $J$
in the interior unitarity class.
In this case the contributions to the observables and to the objective function
$\im \tilde \phi_5$
solely come from the interior unitarity class $I$;
$A_0^I=A_0$, $\im \phi_+^{\,I}=\im \phi_+$, $\Sge^I=\Sge$ and
$\im \tilde \phi_5^I= \im \tilde \phi_5$.
The normalized dimensionless total cross section is reconstructed
by substituting Equation~(\ref{eq:ak1}) into
\be
A_0=
\sum_{J=0}^{M_1}
J \left(
a_{0}^J+a_{1}^J+ a_{11}^J+a_{22}^J
\right)\,
\ee
to give
\be
A_0=
8 \sum_{J=0}^{M_1}
J \left[  r_1 + r_2\left( 1 - \frac{\zeta}{4}\,J^2\right)\right]  \,.
\ee
The Euler-Maclaurin expansion~\cite{eulermac1} for large $J$
is used to write the normalized dimensionless total cross section $A_0$
as an integration over $J$:
\bea
A_0&\approx&
8
\int_0^{M_1} dJ
\left(( r_1 + r_2)\,J - r_2 \frac{\zeta}{4}\, J^3 \right)\\
&\approx&
\frac{M_1^2}{2}
\left\{8 (r_1 +  r_2) - r_2 \zeta M_1^2 \right\}\,.
\lb{eq:A03cI}
\eea

Similarly the imaginary spin average helicity non-flip amplitude
$\im \phi_+^{\,I}$
is reconstructed by
substituting Equation~(\ref{eq:ak1}) into
\be
\im \phi_+=
\sum_{J=0}^{M_1}
J \left\{
a_{0}^J+a_{1}^J+ a_{11}^J+a_{22}^J
\right\}\,
\left( 1 - \frac{\zeta}{4}\,J^2\right)
\ee
to give
\bea
\im \phi_+&\approx&
4 
\int_0^{M_1} dJ
\left\{
(r_1 + r_2)\,J - (2 r_2 + r_1)\frac{\zeta}{4}\,J^3 + r_2
\frac{\zeta^2}{16} J^5
\right\}\\
&\approx&
M_1^2
\left\{ 2 (r_1 +r_2) - (2 r_2 + r_1) \frac{\zeta}{4} M_1^2
+ \frac{r_2 \zeta^2}{24} M_1^4\right\}\,.
\lb{eq:phi+3cI}
\eea
\paragraph{}
The dimensionless normalized elastic cross section $\Sge^I$,
by
substituting Equations~(\ref{eq:ak1}) and (\ref{eq:ak21}) into 
\be
\Sge= 2 \sum_{J=0}^{M_1} J\,
\left( a_{0}^J\,{}^2 + a_{1}^J\,{}^2 + a_{11}^J\,{}^2 + a_{22}^J\,{}^2 +
2 a_{21}^J\,{}^2 \right)\,,
\ee
is reconstructed:
\bea
\Sge &\approx&
\left\{
4 (r_1+r_2)^2 M_1^2 - (r_1+r_2) r_2 \zeta \,M_1^4 +
\frac{r_2^2 \zeta^2}{12} \,M_1^6
\right\}
\nonumber \\
& &\hspace{15mm}+
\frac{M_1^4}{64 \beta^2}
\left\{
1  - \frac{\zeta}{6} \,M_1^2 + \frac{\zeta^2}{128} \,M_1^4
\right\}
\lb{eq:el3cI}
\eea
The modified imaginary single-flip amplitude $\im \tilde \phi_5$ is
reconstructed by substituting Equation~(\ref{eq:ak21}) into
\be
\im {\tilde\phi_5} =
\sum_{J=0}^{M_1}J^2 \,
\left( 1 - \frac{\zeta}{8}\,J^2 \right)\,
a_{21}^J
\ee
leading to
\be
\im {\tilde\phi_5} =
\frac{1}{8 \beta} \sum_{J=0}^{M_1}J^3 \,
\left( 1 - \frac{\zeta}{8}\,J^2 \right)^2\,.
\ee
For large $J$ the modified imaginary single-flip amplitude is written as
\be
\im {\tilde\phi_5} \approx
\frac{1}{8\beta}\,\frac{M_1^4}{4}
\left(1 -\frac{\zeta}{6}M_1^2 + \frac{\zeta^2}{128}M_1^4 \right)\,.
\lb{eq:f52cI}
\ee
An expression for the equality multiplier $\beta$ is found by solving
Equation~(\ref{eq:el3cI}):
\be
\beta =
\frac{M_1^2
\left\{1 -\frac{\zeta}{6}M_1^2 + \frac{\zeta^2}{128}M_1^4 \right\}^{1/2} }
{8
\left\{
\Sge -
\left(
4 (r_1+r_2)^2 M_1^2 - (r_1+r_2) r_2 \zeta \,M_1^4 +
\frac{r_2^2 \zeta^2}{12} \,M_1^6
\right)
\right\}^{1/2}
}\,.
\ee
Rewriting the modified imaginary single-flip amplitude one obtains,
\bea
\im {\tilde\phi_5} &\approx&
\left\{
\Sge -
\left(
4 (r_1+r_2)^2 M_1^2 - (r_1+r_2) r_2 \zeta \,M_1^4 +
\frac{r_2^2 \zeta^2}{12} \,M_1^6
\right)
\right\}^{1/2}
\nonumber \\
& & \hspace{12mm}\times
\frac{M_1^2}{4}
\left\{1 -\frac{\zeta}{6}M_1^2 + \frac{\zeta^2}{128}M_1^4 \right\}^{1/2}\,.
\lb{eq:phi53cI}
\eea
The equality multipliers, $r_1$ and $r_2$, are found by solving
Equations~(\ref{eq:A03cI}) and (\ref{eq:phi+3cI}). The solutions are given
by
\be
r_1 = \frac{A_0^3 \,\zeta \,\left( 1 - 3 \im \phi_+/A_0\right)}
{36 \left(1 - 2 \im \phi_+/A_0\right)^2}
\ee
and
\be
r_2 = \frac{A_0^2 \,\zeta}
{72 \left(1 - 2 \im \phi_+/A_0\right)^2}
\ee
where $\zeta = -t/k^2$.
The equality multiplier $\beta$, with solutions for $r_1$ and $r_2$, is
expressed as
\be
\beta=
\frac{9 (A_0 - 2 \im \phi_+)\,
\sqrt{1 - 2 \im \phi_+/A_0 + 36 \im \phi_+^2/A_0^2}}
{2 A_0 \,\zeta  \sqrt{72 \Sge - 2 A_0^2\,\zeta/(1 - 2 \im \phi_+/A_0)}}\,.
\ee
The optimized modified imaginary single-flip amplitude,
expressed as a function of $r_1$, $r_2$ and $\beta$, becomes
\be
\im \tilde \phi_5=
\frac{(A_0 - 2 \im \phi_+)}{4 A_0\, \zeta}
\frac{\sqrt{1/2 - 2 (1 - \im \phi_+/A_0)\,\im \phi_+}}
{\sqrt{36 \Sge - A_0^2\,\zeta/(1 - 2 \im \phi_+/A_0)}}
\ee
with
\be
J_{\mbox{\footnotesize{max}}} = \frac{12}{\zeta}
\left( 1 - 2\,\frac{\im \phi_+}{A_0}\right)\,.
\ee
For low momentum transfers the imaginary spin average non-flip amplitude
$\im \phi_+$, expanded to order $t$, is written as
\be
\im \phi_+ \approx
\frac{A_0}{2} \left( 1 + g t \right)\,.
\ee
Under this approximation the maximum $J$
inside the interior unitarity class is independent of $t$ and in the limit
$t \rightarrow 0$, the number of partial waves is finite where
\be
J_{\mbox{\footnotesize{max}}}= \sqrt{ 12 g} \,\,k\,. 
\ee
The equality multipliers in the low $t$ limit become
\be
r_1= \frac{A_0}{72 g^2 k^2}\,\left(\frac{1 + 3 g t}{t}\right)\,,
\ee
\be
r_2=- \frac{A_0}{72 g^2 k^2\,t}
\ee
and
\be
\beta=\frac{\sqrt{2}\, 9 g k^2}{ \sqrt{72 \Sge - 2 A_0^2/(g k^2)}}\,
\left( 1 + g t(2 + 9 g t/8) \right)^{1/2}\,.
\ee
The upper bound on $|\im r_5|$, where
$|\im r_5| = m\, |\im \tilde \phi_5|/(k \,\im \phi_+)$, can be expressed
analytically:
\be
|\im r_5| \leq
\frac{ \sqrt{2} m\,k g}{A_0}\, \sqrt{ 18 \Sge - \frac{A_0^2}{2 g k^2} }\,
\times h(t)
\ee
where
\be
h(t)= \frac{\left( 1 + g t(2 + 9 g t/8) \right)^{1/2}}{(1 + gt)}\,.
\ee
The variable $h(t)$ is finite at $t=0$ and changes \lq slowly\rq
$\;$over the CNI region.
Writing $A_0= k^2 \sgt/\pi$ and $\Sge= k^2 \sge/\pi$, enables the bound on
$|\im r_5|$ to be expressed as
\be
|\im r_5|  \leq
m\, \sqrt{g}\,
\left(\frac{36 \pi g \,\sge}{\sgt^2} -1 \right)^{1/2}\times h(t)\,.
\ee
\subsection{Results}
The value of the bound on $|\im r_5|$ is given in Tables~\ref{tab:r54c001}
and \ref{tab:r54c01} with the values of the equality multipliers.
The most noticeable feature of the bound is its size at
low momentum transfers, having a value of $0.89$ at $\sqrt s=52.8$~GeV,
$t=-0.001~(\mbox{GeV}/c)^2$.
The optimized partial waves, at $\sqrt s =52.8$~GeV and
$t=-0.001~(\mbox{GeV}/c)^2$, 
is shown in Figure~\ref{fig:ak1_4c}.
The partial wave series terminates at $J=231$ which is the
upper $J$ limit, $M_1$, for the interior unitarity class $I$.
When considering both the interior and boundary unitarity classes,
values of $J > M_1 $ are permitted.
\begin{table}[!h]
\textwidth 70mm
\begin{center}
\setlength{\tabcolsep}{1.1pc}
\caption
{$|\im r_5|^{\mbox\small{max}}$ inside the interior region at
$t=-0.001~(\mbox{GeV}/c)^2$.}
\label{tab:r54c001}
\begin{tabular}{cccccc}
\hline \hline
$\sqrt s~(\mbox{GeV})$ &
      $r_1$ &
      $r_2$ &
       $\beta$ &
      $J_{\mbox{\footnotesize{max}}}$ &
        $|\im r_5|$\\
\hline
19.4 &
$-12.54$&
12.77&
90&
81&
0.97\\
 23.5   &
 $-12.56$&
 12.79&
 117&
 98&
 0.92\\
30.7 &
$-12.02$&
12.24&
158&
131&
0.92\\
44.7&
 $-11.22$&
 11.44&
 217&
 195&
 1.05\\
52.8&
 $-11.53$&
 11.76&
 293&
 231&
 0.89\\
 62.5&
$-11.56$&
11.79&
358&
276&
0.86\\
  \hline   \hline
\end{tabular}
\end{center}
\end{table}
\textwidth 160mm
\begin{table}[!h]
\textwidth 70mm
\begin{center}
\setlength{\tabcolsep}{1.1pc}
\caption
{$|\im r_5|^{\mbox\small{max}}$ inside the interior region at $t=-0.01~(\mbox{GeV}/c)^2$.}
\label{tab:r54c01}
\begin{tabular}{cccccc}
\hline    \hline
$\sqrt s~(\mbox{GeV})$ &
      $r_1$ &
      $r_2$ &
       $\beta$ &
      $J_{\mbox{\footnotesize{max}}}$ &
        $|\im r_5|$\\
\hline
19.4 &
$-1.25$&
1.49&
85&
81&
0.97\\
 23.5   &
 $-1.05$&
 1.27&
 111&
 98&
 0.91\\
30.7 &
$-1.00$&
1.22&
150&
131&
0.92\\
44.7&
 $-0.92$&
 1.14&
 204&
 195&
 1.05\\
52.8&
 $-0.95$&
 1.17&
 276&
 231&
 0.89\\
 62.5&
$-0.94$&
1.18&
337&
276&
0.86\\
  \hline  \hline
\end{tabular}
\end{center}
\end{table}
\textwidth 160mm

The bound on $|\im r_5|$, under the approximation
\be
g \approx \frac{\sgt^2}{32 \pi \sge}\,, 
\ee
with momentum transfers in the CNI region is expressed as
\be
|\im r_5|  \leq
m\, \sqrt{\frac{g}{8}}\times h(t)
\ee
and in the zero momentum transfer limit, $t \rightarrow 0$, the bound on
$|\im r_5|$ is finite and can be expressed analytically as
\be
|\im r_5| \leq m\,\sqrt{\frac{g}{8}}\,.
\ee
This approximation generates a \lq stricter\rq  $\;$ bound on $|\im r_5|$.
The results are given in 
Table~\ref{tab:gbnb}.
\begin{table}[!hbt]
\textwidth 70mm
\begin{center}
\setlength{\tabcolsep}{0.5pc}
\caption
{$|\im r_5|^{\mbox\small{max}}$, with an
approximation for $g$, over the CNI region.}
\label{tab:gbnb}
\vspace*{3mm}
\begin{tabular}{cccc}
\hline\hline
$\sqrt s~(\mbox{GeV})$ &
       $t= 0~(\mbox{GeV}/c)^2$&
       $t=-0.001~(\mbox{GeV}/c)^2$&
       $t=-0.01~(\mbox{GeV}/c)^2$\\
\hline
19.4 &
0.803&
0.805&
0.825\\
 23.5   &
 0.805&    0.808&      0.827\\
30.7 &  0.819&     0.821&     0.842\\
44.7& 0.839&     0.841&     0.864\\
52.8& 0.841&     0.843&     0.866\\
 62.5& 0.846&     0.848&     0.871\\
  \hline\hline
\end{tabular}
\end{center}
\end{table}
\textwidth 160mm
\begin{figure}[!h]
\begin{center}
\mbox{\epsfxsize 55mm
      \epsfbox{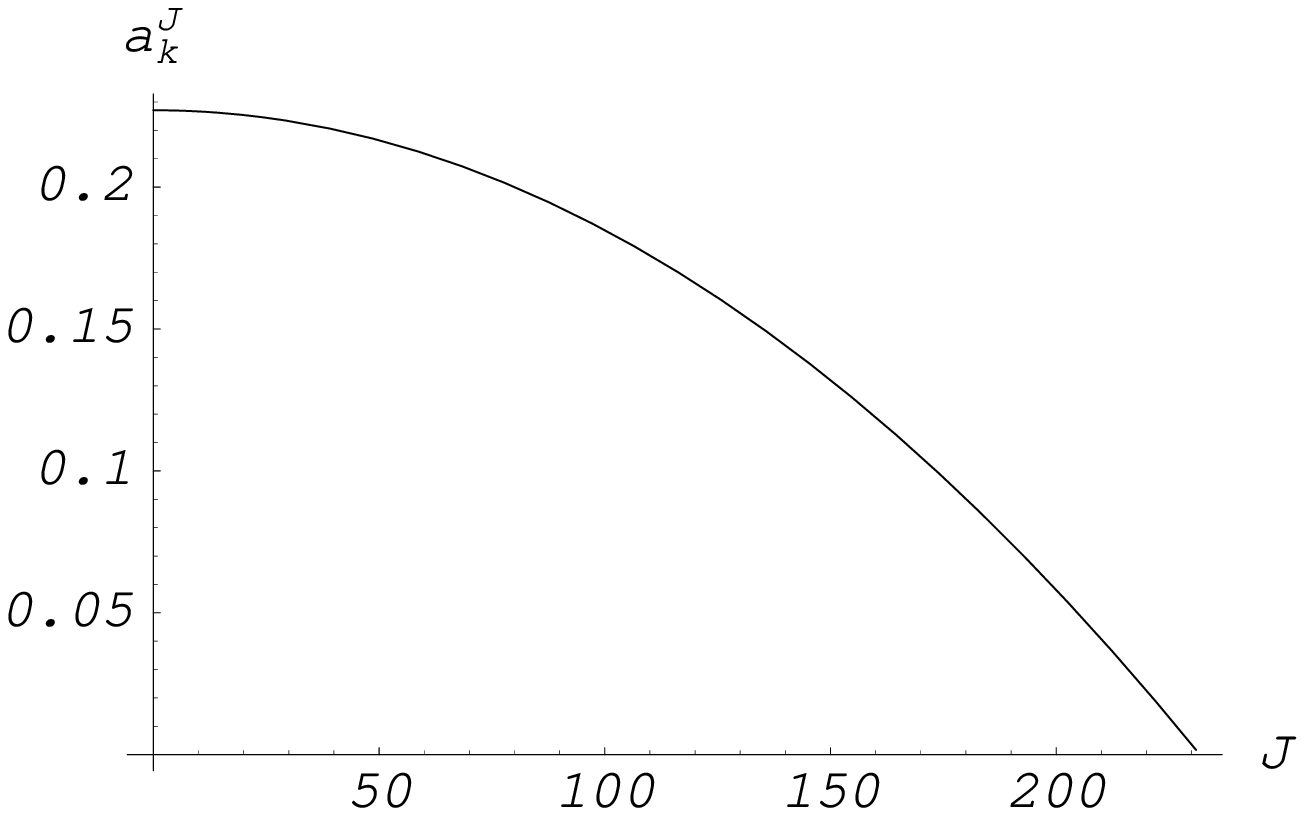}}
\mbox{\epsfxsize 55mm
      \epsfbox{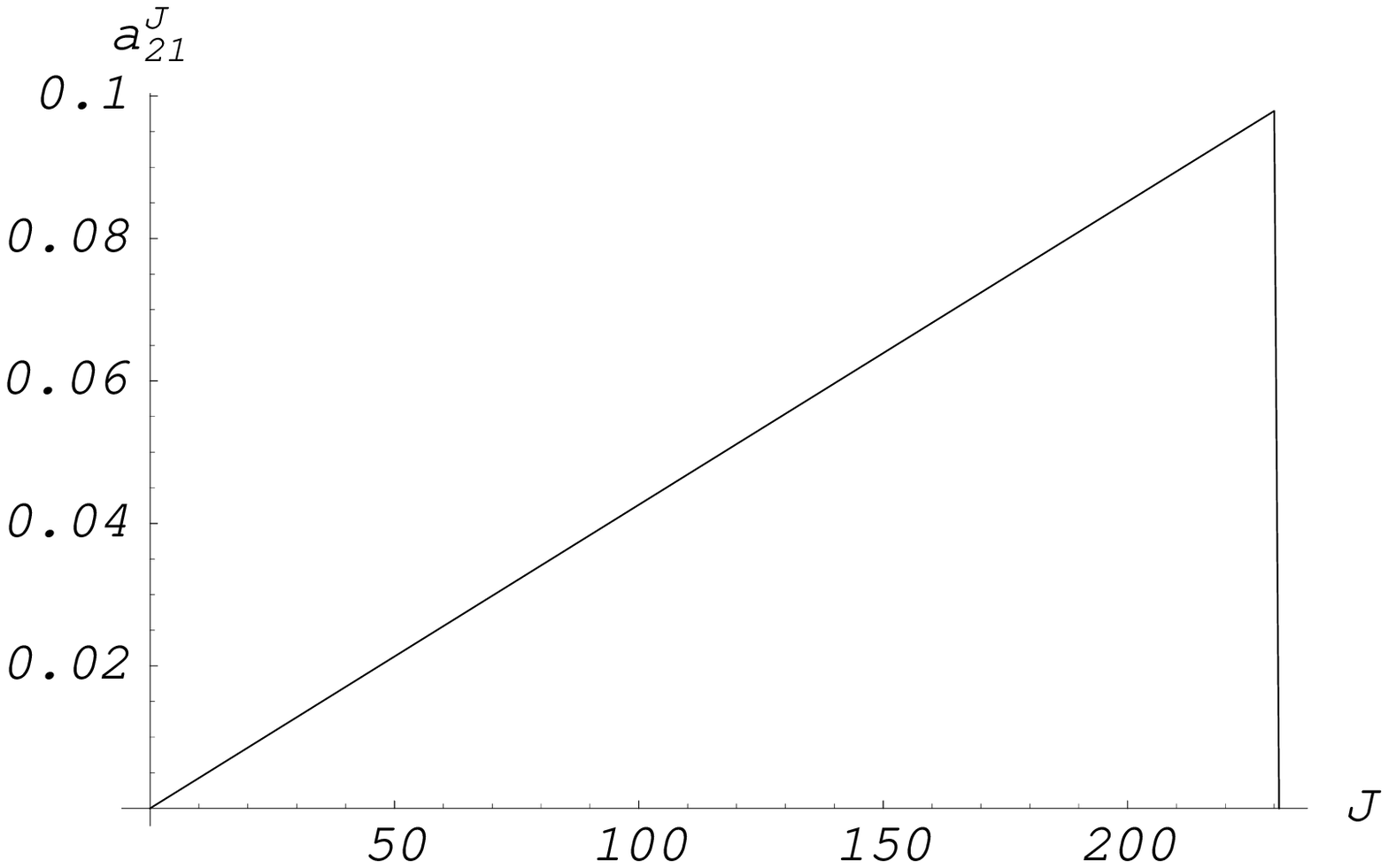}}
\vspace{-5mm}
\end{center}
\caption{$a_{k}^J\;(k=0, 1, 11, 22)$ and $a_{21}^J$ in the interior class;
$\sqrt s = 52.8~\mbox{GeV}$,
$t=-0.001~(\mbox{GeV}/c)^2$.\lb{fig:ak1_4c}}
\end{figure}
\subsection{Solution of Interior and Boundary Classes}
Consider the set of classes
$I \cup B \equiv
I^W \cup I^X \cup B^W \cup B^X$.
The boundary unitarity classes are
\be
B^{X}  \equiv \left\{J | a_{11}^J - a_{11}^J{}^2- a_{21}^J{}^2 =0,\;
M_1+1 \leq J \leq M_2\right\}
\ee
and
\be
B^{W_0}  \equiv \left\{J | a_0^J = 0,\; M_1+1 \leq J \leq M_2\right\}
\ee
where $M_1={\tt Floor}[\sqrt{4/\zeta(1 + r_1/r_2)}\,]$,
$M_2={\tt Floor}[\sqrt{8/\zeta}\,]$ and $\zeta=-t/k^2$.
The contribution to $|\im r_5|$ from the boundary unitarity class $B$ can
range from $0\%$ to $100\%$ and the contribution to $|\im r_5|$, from the boundary unitarity class, can be
selected without violating any of the constraints and this
contribution can be made arbitrarily small.
\subparagraph{}
Consider the case with $\Sge^B= 0.1  \Sge$, $\Sge^I=0.9  \Sge$,
at $\sqrt s = 52.8~\mbox{GeV}$ and 
$t=-0.001~(\mbox{GeV}/c)^2$.
The maximum contribution to $|\im r_5|$ is 34.7 where
$|\im r_5^I| \leq 0.5$ and
$|\im r_5^B| \leq 34.2$.
The case with $\Sge^B= 0.01  \Sge$, $\Sge^I=0.99  \Sge$,
leads to $|\im r_5| \leq 11.6$ where
$|\im r_5^I| \leq 0.8$ and
$|\im r_5^B| \leq 10.6$.
The bound on $|\im r_5^B|$ falls when the fraction of $\Sge$ in the
boundary unitarity class is reduced. The partial wave amplitudes in this
region also become smaller in amplitude and contribute less to the bound on
$|\im r_5^B|$.
The fraction of $\Sge$ in the boundary unitarity class can be further reduced
until the contribution from this class to $|\im r_5|$ is negligible
in comparison with the contribution from the interior unitarity class. In
this limit the bound is, as before, 
\be
|\im r_5|  
\leq
m\, \sqrt{g}\,
\left(\frac{36 \pi g \,\sge}{\sgt^2} -1 \right)^{1/2}\times h(t)
\ee
or, under the approximation $g \approx \sgt^2/(32 \pi \sge)$,
\be
|\im r_5|  \leq
m \sqrt{\frac{g}{8}}\times h(t)
\ee
where
\be
h(t)= \frac{\left( 1 + g t(2 + 9/8\, g t) \right)^{1/2}}{(1 + gt)}\,.
\ee
The bound is identical to the bound when only the interior unitarity class is
considered.
A finite number of partial waves at low momentum transfer
ensures a finite value for the sum
\be
\im \tilde\phi_5 = \sum_J
 J^2\,a_{21}^J  \,
\left(1 - \frac{\zeta}{8}\,J^2\right)
\ee
and consequently a finite upper bound on $|\im r_5|$ of less than unity.
\section{Error on Bound}
\setcounter{equation}{0}
The upper bound on the imaginary single helicity-flip amplitude, modified by
a kinematical factor at zero momentum transfer, is given by
\be
|\im r_5| \leq  m \sqrt{\frac{g}{8}}\, \left(\frac{36 \pi g \sge}{\sgt^2}-1\right)^{1/2}
\,.
\lb{eq:r51}
\ee
There are errors on all the experimental quantities in Eqn.~(\ref{eq:r51})
and consequently the upper bound on $|\im r_5|$ has an uncertainty.
The experimental quantities $g$, $\sgt$ and $\sge$ have a nominal value plus
an uncertainty: $g\pm \Delta g$, $\sgt \pm \Delta \sgt$ and
$\sge \pm \Delta \sge$.
What are the values of $\Delta g$, $\Delta \sgt$ and $\Delta \sge$?
Consider the value of $\im r_5$ at $\sqrt s = 52.8$~GeV;
$|\im r5| \leq 0.891$ with
$g =6.435\pm 0.14\mbox{GeV}^{-2}$~\cite{pereira},
$\sgt = 42.906$~mb~\cite{pdb} and $\sge = 7.407$~mb~\cite{pdb}. 
The value of $\Delta g$ is known but we must calculate $\Delta \sgt$ and
$\Delta \sge$.
\paragraph{}
A parameterization for the total and elastic cross section in elastic {\it pp}
collisions~\cite{pdb} allows a value for the cross sections to be found and
a value of their uncertainties to be calculated. 
Each cross section is parameterized as
\be
\sigma(p)= A + Bp^n + C \log^2(p) + D \log(p)
\ee
where $\sigma$ is in mb and $p$ is the laboratory momentum in GeV/$c$.
The uncertainty in $\sigma$ is given by
\clearpage
\bea
\Delta \sigma &=&
\left\{
\left(\frac{\partial \sigma}{\partial A}\right)^2 (\Delta A)^2
+ \left(\frac{\partial \sigma}{\partial B}\right)^2 (\Delta B)^2
+\left(\frac{\partial \sigma}{\partial n}\right)^2 (\Delta n)^2 +
 \right. \nonumber \\
& & \hspace*{10mm}\left.
\left(\frac{\partial \sigma}{\partial C}\right)^2 (\Delta C)^2
+ \left(\frac{\partial \sigma}{\partial D}\right)^2 (\Delta D)^2
\right\}^{1/2}\,.
\lb{eq:unc}
\eea
The fitted parameters $A$, $B$, $n$, $C$ and $D$ are given in
Table~\ref{tab:fits}.
\begin{table}[!h]
\textwidth 50mm
\begin{center}
\setlength{\tabcolsep}{0.5pc}
\caption{Fitted parameters for {\it pp} scattering\label{tab:fits}}
\begin{tabular}{cccccc}
\hline
\hline
Reaction&
      $A$ &
       $B$ &
      $n$&
        $C$&
        $D$\\
\hline
$\sgt$&
$48.0\pm0.1$&
$-$&
$-$&
$0.522\pm0.005$&
$-4.51 \pm0.05$ \\
$\sge$&
$11.9\pm0.8$&
$26.9\pm1.7$&
$-1.21\pm0.11$&
$0.169\pm0.021$&
$-1.85 \pm0.26$ \\
\hline
\hline
\end{tabular}
\end{center}
\end{table}
\textwidth 160mm
\newline
Using Eqn.~(\ref{eq:unc}), and the values of the fitted parameters,
the uncertainties $\Delta \sgt$ and $\Delta \sge$ can be written as
\be
\Delta \sgt =
\sqrt{
0.01 + 2.5\times 10^{-5} \log^4(p) + 2.5 \times 10^{-3} \log^2(p)
}
\ee
and
\be
\Delta \sge =
\sqrt{
0.64 + 2.89 p^{-2.42} + 12.819 p^{-4.42} +
4.41\times 10^{-4} \log^4(p) + 0.0676\log^2(p)
}\,.
\ee
A laboratory beam momentum of $p = k \sqrt s /m =1485$~GeV/$c$ at $\sqrt s =52.8$~GeV gives
$\Delta \sgt=0.463~\mbox{mb}$, or $1.08\%$ of $\sgt$ and
$\Delta \sge=2.345~\mbox{mb}$, or $31.66\%$ of $\sge$.
\paragraph{}
The uncertainty in $\im r_5$ is
\be
\Delta \im r_5 =
\sqrt{
\left(\frac{\partial \im r_5}{\partial g}\right)^2 (\Delta g)^2
+ \left(\frac{\partial \im r_5}{\partial \sgt}\right)^2 (\Delta \sgt)^2
+\left(\frac{\partial \im r_5}{\partial \sge}\right)^2 (\Delta \sge)^2 
}\,.
\lb{eq:uncr5}
\ee
At $\sqrt s =52.8$~GeV, 
$g =6.435\pm 0.14\mbox{GeV}^{-2}$,
$\sgt = 42.906 \pm 0.463 $~mb and $\sge = 7.407 \pm 2.345$~mb.
The uncertainty $\Delta \im r_5 = 0.049$ and the upper bound on
$|\im r_5|$ is  $0.891 \pm 0.049$.
\paragraph{}
The approximation
\be
g \approx \frac{\sgt^2}{32 \pi \sge}
\ee
can be used to write the bound on $|\im r_5|$  as
\be
|\im r_5| \leq m \sqrt{\frac{g}{8}}
\ee
at zero momentum transfer.
The uncertainty $\Delta \im r_5$ is simply
\be
\Delta \im r_5 =
\sqrt{
\left(\frac{\partial \im r_5}{\partial g}\right)^2 (\Delta g)^2
} \,.
\lb{eq:uncr5g}
\ee 
At $\sqrt s =52.8$~GeV, 
$g =6.435\pm 0.14\mbox{GeV}^{-2}$, $|\im r_5| \leq 0.846$ and
$\Delta \im r_5=0.009$ or $|\im r_5| \leq 0.846 \pm 0.009$.
The error on $\sge$ being large has relatively little effect on the uncertainty
of the bound.
\section{Conclusion}
A bound on the size of the imaginary part of the single helicity-flip amplitude
, in the CNI region, was obtained using the Lagrange variational method of 
Einhorn and Blankenbecler with $\sgt$, $\sge$ and diffraction slope expressed 
as equality constraints and unitarity expressed as two inequality constraints.
An important feature associated with the bound is the number of partial waves at low momentum
transfer. In the CNI region the number of waves is finite, that is, there is
no singularity in $\sqrt{-t}$ as $t \rightarrow 0$. 
This feature ensures that a useful bound exists and, in fact,
the bound limits $|\im r_5|$ to values less than unity near $t=0$.
With more constraints in the system an improved bound could be obtained 
although any additional constraints must be sufficiently well 
known experimentally
and be computationally tractable.
As the bound of 0.84 on the helicity flip amplitude ratio
is less than $(\mu_p - 1)/2 =0.896$ at the high energies considered,
the coefficient of $(t_c/t)$ in the expression for the asymmetry
is constrained to be positive and therefore the analyzing power
is positive for at least a significant part of the interference region.
Though the bound is not sufficient to limit the polarization
error to the recommended $5\%$, it does permit the use of
proton proton elastic collisions in the CNI region as a
relative polarimeter.
\setcounter{equation}{0}

\end{document}